\documentstyle[prl,aps,epsfig,twocolumn,epsf]{revtex}

\renewcommand{\vec}[1]{\mbox{\boldmath$#1$}}

\begin{document}

\twocolumn[\hsize\textwidth\columnwidth\hsize\csname  
@twocolumnfalse\endcsname

\title{Dephasing of a qubit coupled with a point-contact detector}

\author{Y. Rikitake, H. Imamura, Y. Utsumi, M. Hayashi, H. Ebisawa}

\address{Graduate School of Information Sciences, Tohoku University,
  Sendai 980-8579, Japan
  }

\maketitle


\begin{abstract}
  The dephasing of a qubit coupled with 
  a point-contact detector is theoretically studied.
  We calculate the time evolution of the reduced density matrix of
  qubit by using the perturbation expansion.  
  We show that the  dephasing rate is proportional to the temperature
  at zero bias-voltage, while it is  proportional to the bias-voltage
  when the bias-voltage is large.  We also evaluate the dephasing rate
  by using the real time renormalization group method and show that
  the higher order processes of the particle-hole excitation enhances
  dephasing of qubit.
\end{abstract}

\vskip1pc]


Much attention has been devoted in recent years to quantum computation
and quantum information,  which are the new technology of
information processing based on quantum mechanical
principles\cite{nielsen_book}.
The dephasing\cite{weis_book} of a qubit is one of the fundamental problems 
in quantum information sciences.  Gruvitz has studied the dephasing
and collapse of a qubit in a double-dot(DD) caused by the 
continuous measurement by a point-contact(PC)
detector\cite{gurvitz1997,gurvitz1998}.
He showed that the collapse and the role of the observer in quantum
mechanics can be resolved experimentally via a non-destructive
continuous monitoring of a single quantum system. However, extensive
theoretical studies on the dephasing of qubit, such as the temperature
dependence and bias-voltage dependence, are needed to construct
the reliable quantum information processing system.

In this paper,  we study the dephasing of a qubit coupled with 
a PC detector.
Using the lowest order approximation, we show that the dephasing rate
is proportional to the temperature at zero bias-voltage while it is
proportional to the bias-voltage in the limit of large bias-voltage.  
We also evaluate the dephasing rate by using the real time
renormalization group method developed by
Shoeller\cite{shoeller2000,keil2000} 
and show that the dephasing rate is enhanced by the higher
order processes of the particle-hole excitation.
  
%
%

\begin{figure}[h]
  \epsfxsize=0.9\columnwidth 
  \centerline{\hbox{
      \epsffile{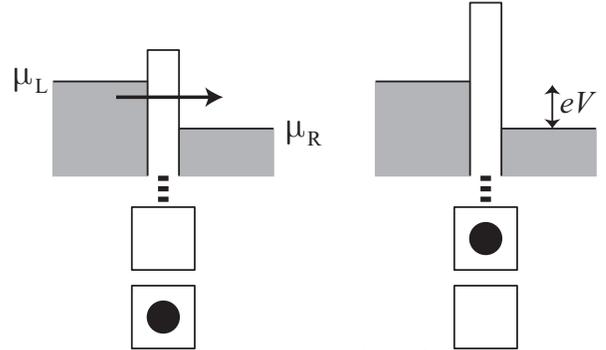}}}
\caption{The double-dot (qubit) coupled with a point-contact detector.
  $\mu_{L}$  and $\mu_{R}$ are the chemical potential of the left and
  right reservoirs  of point-contact and $eV$ is the difference
  between them.  The filled circle represents the electron in a double-dot.
  The height of the potential barrier of the point-contact is modified
  by the electron state of the double-dot.  The current through the
  point contact flows only when the electron  occupies the lower dot.}
  \label{fig:model}
\end{figure}

The system we consider is a DD coupled with a PC
detector.  The PC is placed near the upper dot as shown in
Fig. \ref{fig:model}\cite{gurvitz1997,gurvitz1998}.
The barrier height of the PC, therefore the tunneling current through
 PC, is modified by the electron state of DD.
We assume that current can flow if and only if the electron in the DD
occupies the lower dot.  In our system, the dephasing of qubit is 
caused by the interaction between the qubit and the PC.

%
%

We can map the electron state of the DD into that of the equivalent 
two-level system and the Hamiltonian of the DD can be expressed
in terms of Pauli spin matrices $\vec{\sigma}$.
The electron state with upper(lower) dot occupied
corresponds to the eigenstates of $\sigma_z$ 
$|\uparrow\rangle$ $(\sigma_z = 1)$
($|\downarrow\rangle$ $(\sigma_z=-1)$).
The total Hamiltonian of the system is given by 
\begin{eqnarray}
  H &=& H_{qb}+H_L+H_R+H_{int},\label{eq:hamiltonian}
\end{eqnarray}

where

\begin{eqnarray}
  H_{qb} &=& \frac{\epsilon_{_{C}}}{2} \sigma_z, \\
  H_L &=& \sum_l \epsilon_l c_l^{\dag} c_l,  \ \ H_R = \sum_r \epsilon_r c_r^{\dag} c_r, \\
  H_{int} &=& \frac{1}{2} (\sigma_z + 1)
  \sum_{l,r}(\Omega c_l^{\dag} c_r +\rm{h.c.}).
\end{eqnarray}
Here $H_{qb}$, $H_{L(R)}$ and $H_{int}$ are the Hamiltonians
of the qubit, the left(right) reservoir
of PC and qubit-PC interaction, respectively.
$\epsilon_{l(r)}$ are the energy levels in the left(right) reservoir,
and $\Omega$ is the tunneling matrix element of the PC.
When the upper dot is occupied $(\sigma_{z}=-1)$,
the tunneling current does not flow through the PC.
Since our interest is in the dephasing of the qubit,
we neglect the tunneling between upper and lower dots.

We consider the following reduced density matrix of the qubit:
\begin{eqnarray}
  \rho = Tr_{\rm PC}\ \rho_{\rm tot} =
  \left(
  \begin{array}{cc}
    \rho_{\uparrow\uparrow} & \rho_{\uparrow\downarrow} \\
    \rho_{\downarrow\uparrow} & \rho_{\downarrow\downarrow}
  \end{array}
  \right),
  \label{eq:rho}
\end{eqnarray}
where $\rho_{\rm tot}$ is the density matrix of the total system
consisting of the qubit and PC, $Tr_{\rm PC}$ denotes tracing out the degrees
of reservoirs.

The time evolution of the total density matrix 
$\rho_{\rm tot}(t)$ is described by the von Neumann equation, $ i  \hbar
\dot{\rho}=[H,\rho]$.
Due to interaction with the detector, the time evolution of 
the reduced density matrix $\rho(t)$ is not unitary and 
it approaches the statistical mixture
represented by the diagonal matrix in the limit of $t\rightarrow\infty$. 
Therefore, we can evaluate the dephasing of the qubit by 
calculating the time evolution of the off-diagonal element of reduced
density matrix $\rho_{\uparrow\downarrow}(t)$.

We assume that the density matrix of the total system at the initial
time $t=0$ is in product form and the qubit and reservoirs of PC are
decoupled, and the reservoirs are in thermal
equilibrium\cite{weis_book,feynman1963}.  We then have
\begin{eqnarray}
  \rho_{{\rm tot}}(0)=\rho(0)\otimes\rho_{L0}\otimes\rho_{R0},
  \label{eq:init}
\end{eqnarray}
where
\begin{eqnarray}
  \rho_{{ L}0}&=&\frac{{\rm e}^{-\sum_l(\epsilon_l-\mu_{\rm L})c_l^{\dag} c_l/k_{\rm B}T}}
  {{\rm Tr}_{ L}{\rm e}^{-\sum_l(\epsilon_l-\mu_{\rm L})c_l^{\dag} c_l/k_{\rm B}T}}, \\
  \rho_{{ R}0}&=&\frac{{\rm e}^{-\sum_r(\epsilon_r-\mu_{\rm R})c_r^{\dag} c_r/k_{\rm B}T}}
  {{\rm Tr}_{ R}{\rm e}^{-\sum_r(\epsilon_r-\mu_{\rm R})c_r^{\dag} c_r/k_{\rm B}T}}. 
\end{eqnarray}

The time evolution of the off-diagonal element $\rho_{\uparrow\downarrow}(t)$
obeys the generalized master equation defined by
\begin{eqnarray}
  \!\!  \left[\frac{{\rm d}}{{\rm d} t}+ i
    \frac{\epsilon_{_{C}}}{\hbar}\right]
  \!\rho_{\uparrow\downarrow}(t)
  \!\!=\!\!\!\! \int_0^t \!\!\!{\rm d}
  t^{\prime}\rho_{\uparrow\downarrow}(t^{\prime})\Sigma(t \!-\! t^{\prime}),
  \label{eq:master_eq}
\end{eqnarray}
where $\Sigma$ is a `self-energy' which describes
the qubit-detector interactions\cite{shon_book,shnirman1998}.

\begin{figure}[h]
  \epsfxsize=0.9\columnwidth 
  \centerline{\hbox{
      \epsffile{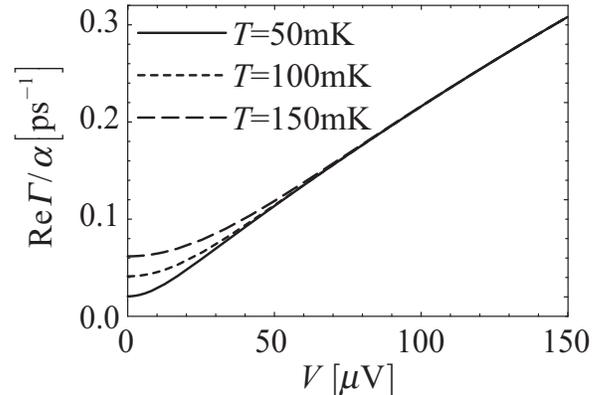}}}
\caption{Dephasing rate of the qubit as a function of 
      bias-voltage $V$ applied to the point contact.
      Solid, dotted and dashed lines represent the dephasing rate for
      $T=50, 100$ and $150$mK, respectively.  The high
      frequency cutoff is assumed to be $\hbar D=20$K.}
  \label{fig:dep}
\end{figure}

We first calculate the self-energy $\Sigma(t - t^{\prime})$ by using
the  perturbation expansion with respect to the interaction
term $H_{int}$.  In the lowest order approximation, the self-energy
is given by
\begin{equation}
  \Sigma(t-t^{\prime})=\exp[-i\epsilon_{_{C}}(t-t^{\prime})]\gamma(t-t^{\prime}),
\end{equation}
where $\gamma(t-t^{\prime})$ is the propagator of 
the particle-hole excitation defined as
\begin{eqnarray}
 &&\mbox{\hspace{-1em}}\gamma(t  -  t^{\prime})
  =
  \alpha  \left(\frac{\pi k_{\rm B}T}{\hbar}\right)^2\nonumber\\
  &&\mbox{\hspace{1em}}\times
  \frac{\cos[eV(t-t^{\prime})/\hbar]}
  {\sinh^2[\pi k_{\rm B}T(t  -  t^{\prime}  -  i /D)/\hbar]}.
  \label{eq:p-h}
\end{eqnarray}
Here $D$ is the high frequency cutoff and $\alpha$ is the dimensionless
conductance of the PC defined as $\alpha = 2\Omega^{2}N_{L}N_{R}$,
where $N_{L(R)}$ is the density of states in the left(right) reservoir.
Suppose that the reduced density matrix varies
very slowly compared with the time scale of the life time of
the particle-hole excitation $\hbar/ k_{\rm B}T$, 
$\rho_{\downarrow\uparrow}(t)$ is written as
\begin{equation}
  \rho_{\downarrow\uparrow}(t) =
  \rho_{\downarrow\uparrow}(0)
  {\rm e}^{-i\epsilon_{_{C}}t/\hbar}
  {\rm e}^{-\Gamma t},
\end{equation}
where
\begin{equation}
  \Gamma = \int_0^{\infty}{\rm d} t \Sigma(t).
\end{equation}
The dephasing rate, or the decay rate of
$\rho_{\uparrow\downarrow}(t)$,  is given by ${\rm Re}\Gamma$. Since
the dephasing is caused by the particle hole
excitation in reservoirs, it depends on the temperature $T$ and
bias-voltage $V$ as shown in Fig. \ref{fig:dep}.
 
It is easy to show that, in the lowest order approximation for the
self-energy, the dephasing rate at $V=0$ is given by
\begin{eqnarray}
  {\rm Re}\ \Gamma = \alpha \frac{\pi k_{\rm B}T}{\hbar}
  \label{eq:d_m0}
\end{eqnarray}
and is proportional to the temperature $T$ as shown in Fig. \ref{fig:dep}.
This can be explained as follows.  The dephasing rate is determined by
the number of particle-hole excitations in the reservoirs which
interact with the qubit.  The number of particle-hole excitations
at $V=0$ is proportional to the temperature $T$.
On the other hand, the number of particle-hole excitations for $eV\gg
k_{\rm B}T$ is proportional to the bias-voltage $V$.
Indeed, the dephasing rate for $eV\gg
k_{\rm B}T$  is given by 
\begin{eqnarray}
  {\rm Re}\ \Gamma = \alpha \frac{1}{2}\frac{eV}{\hbar}
  \label{eq:d_kT0}
\end{eqnarray}
and is proportional to $V$ as shown in Fig. \ref{fig:dep}.

Next we go beyond the lowest order approximation.
We employ the real time renormalization group (RTRG) method
developed by Schoeller\cite{shoeller2000,keil2000}.
Following Shoeller, we introduce the short-time cutoff
$t_c$ in the propagator of the particle-hole excitation by
\begin{eqnarray}
  \gamma_{t_c}(\tau)=\gamma(\tau)\theta(\tau-t_c),
\end{eqnarray}
where $\theta(t)$ is a step function.  In Laplace space, the self-energy $\Sigma(z)$ is expressed by
\begin{eqnarray}
  \Sigma(z) = \Sigma_{t_c}(z)+{\cal F}_{t_c},
\end{eqnarray}
where $\Sigma_{t_c}(z)$ includes only the time scales which are
precisely smaller than $t_c$ and
${\cal F}_{t_c}$ consisting of the other time scales is the
functional of renormalized correlation function $\gamma_{t_c}$,
vertex $g_{t_c}$ and frequency $\omega_{t_c}$

Let us increase the cutoff by an infinitesimal amount $t_c \rightarrow
t_c + dt_c$.  The change of $\gamma_{t_c}(\tau)$ is given by
\begin{eqnarray}
  d\gamma(\tau)&=&\gamma_{t_c}(\tau)-\gamma_{t_c+dt_{c}}(\tau)\nonumber\\
  &=&\gamma(t_c)\delta(\tau-t_c)dt_c.
\end{eqnarray}
The change of the self-energy
$\Sigma_{t_c + dt_{c}}(z)$ caused by $d\gamma(\tau)$ 
can be expressed by $d \gamma$,
$g_{t_c}$ and $\omega_{t_c}$.  The change of $g_{t_c}$ and
$\omega_{t_c}$ due to $d\gamma(\tau)$ can also be represented by $d\gamma$.
As we increase $t_c$, the long time scale of $\gamma$ is renormalized.
Since $\gamma(\tau)$ is a decreasing function of $\tau$ with the life
time $\hbar/k_{\rm B} T$, ${\cal F}_{t_c}$ is zero for $t_c
\rightarrow \infty$.  Therefore, the self-energy $\Sigma(z)$
in Laplace space is expressed as
\begin{eqnarray}
  \Sigma(z)=\lim_{t_c\rightarrow\infty}\Sigma_{t_c}(z).
\end{eqnarray}

The renormalization group (RG) equations is obtained by calculating the
renormalization of $\omega_{t_c}$, $g_{t_c}$ and $\Sigma_{t_c}$ due to the
infinitesimal deviation $d\gamma_{t_c}$.  The straightforward calculation
leads to the following RG equations:
\begin{eqnarray}
&& \mbox{\hspace{-2.5em}} \frac{{\rm d}\Sigma_{t_c}(z)}{{\rm d}
  t_c}=\!\!\int_0^{\infty} \!{\rm d} t
\! \left(\frac{{\rm d}\gamma_{t_c}}{{\rm d} t_c}\right)\!(t) a_{t_c}(t)b_{t_c}(0)\\
&&  \mbox{\hspace{-2.5em}} \frac{{\rm d}\omega_{t_c}}{{\rm d} t_c}=i
  \int_0^{\infty}{\rm d} t \left(\frac{{\rm d}\gamma_{t_c}}{{\rm d}
  t_c}\right)(t) g_{t_c}(t)g_{t_c}(0)\\
&& \mbox{\hspace{-2.5em}}\frac{{\rm d} g}{{\rm d} t_c}=0 \\
&& \mbox{\hspace{-2.5em}}  \frac{{\rm d} a_{t_c}}{{\rm d}
  t_c}=\int_0^{\infty}{\rm d} t\int_{-\infty}^0{\rm d} t^{\prime}
  \left(\frac{{\rm d}\gamma_{t_c}}{{\rm d}
  t_c}\right)\!(t-t^{\prime})\nonumber\\
&&\mbox{\hspace{-1.5em}}\times\left[a_{t_c}\!(t)g_{t_c}\!(0)g_{t_c}\!(t^{\prime})
  \! -\! g_{t_c}\! (t)a_{t_c}\! (0)g_{t_c}\! (t^{\prime})\right] \\
&&  \mbox{\hspace{-2.5em}} \frac{{\rm d} b_{t_c}}{{\rm d} t_c}=-\int_0^{\infty}{\rm d} t\int_{-\infty}^0{\rm d}
  t^{\prime} \left(\frac{{\rm d}\gamma_{t_c}}{{\rm d}
  t_c}\right)\!(t-t^{\prime})\nonumber\\
&& \mbox{\hspace{0em}}\times g_{t_c}(t)g_{t_c}(0)b_{t_c}(t^{\prime} ),
\end{eqnarray}
where
\begin{equation}
  \left(\frac{{\rm d}\gamma_{t_{c}}}{{\rm d} t_c}\right)\!(\tau)=\gamma(t_c)\delta(\tau-t_c).
\end{equation}

We numerically solve the above RG equations with the initial condition
$\omega=\epsilon_{_{C}}/\hbar$ and  $g=a=b=1$ at $t_{c}=0$.
Once we obtain the self-energy in Laplace space, 
the off-diagonal element of reduced density
matrix in Laplace space is given by
\begin{equation}
  \rho_{\downarrow\uparrow}(z)=\frac{\rho_{\downarrow\uparrow}(0)}{z
  - i \epsilon_{_{C}} - \Sigma(z)}.
\end{equation}
Finally, we obtain $\rho_{\downarrow\uparrow}(t)$ by performing an inverse
Laplace transformation.

\begin{figure}[h]
  \epsfxsize=0.9\columnwidth 
  \centerline{\hbox{
      \epsffile{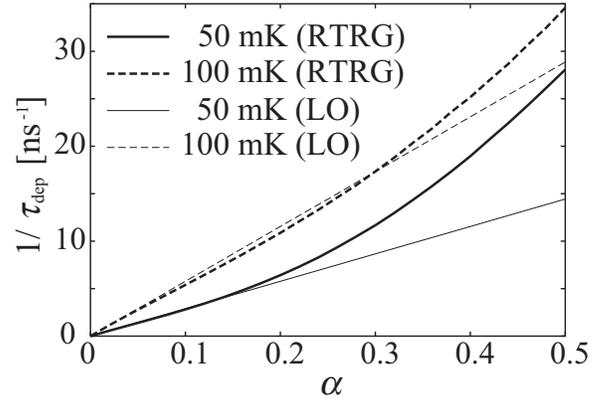}}}
\caption{Dephasing rate calculated by using the real time
  renormalization group (RTRG) method at $k_{\rm B}T=50(100)$mK 
  is plotted by the solid(dotted) line against the dimensionless
  conductance $\alpha$.
  Thin solid(dotted) line represents the dephasing rate obtained by
  using the lowest order (LO) approximation at $k_{\rm
  B}T=50(100)$mK.    The high frequency cutoff is assumed to be $\hbar
  D=4$K.}
  \label{fig:higher}
\end{figure}

We define the dephasing time $\tau_{dep}$ as the time when the
absolute value $|\rho_{\downarrow\uparrow}(t)|$ becomes one-half
of its initial value $|\rho_{\downarrow\uparrow}(0)|$.
In Fig. \ref{fig:higher} we plot the dephasing rate defined as
$1/\tau_{dep}$ at $V=0$ as a function of $\alpha$.
For small $\alpha$, the result agrees  well with that
obtaind by the lowest order approximation.  For large $\alpha
> 0.2$, however, the dephasing rate becomes larger than that
calculated in the lowest order approximation.  As shown in
Fig. \ref{fig:higher}, the higher order processes of the particle-hole
excitation enhances the dephasing rate, since the number of
particle-hole excitations increases due to the quantum fluctuation
described by those higher order processes.
 
%
%

In conclusion, we have studied the dephasing of a double-dot qubit 
coupled with a point-contact detector.  
The time evolution of the reduced density matrix of the qubit is
calculated by using the perturbation expansion.
In the lowest order approximation of the self-energy, we show that the dephasing rate is proportional to the
temperature at $V=0$, while it is proportional to the
bias-voltage at large bias-voltage, $eV\gg k_{\rm B}T$.
The real time renormalization group method is also applied
to evaluate the dephasing rate of the qubit.
We show that the dephasing rate is enhanced by the higher
order processes of the particle-hole excitation.

One of the author H. I. is supported by MEXT, Grant-in-Aid for
Encouragement of Young Scientists, 13740197, 2001.

\end{document}